\documentclass[twoside,slac_one]{revtex4}
\usepackage{graphicx}
\usepackage{fancyhdr}
\usepackage{amsmath} 
\usepackage{bm}
\usepackage{amsxtra}
\usepackage{amssymb}
\usepackage{amsthm}
\usepackage{latexsym}
\usepackage{lscape}
\usepackage{subfigure}

\def\bs{${B_s^0}$}
\def\bsdec{${B_s^0 \rightarrow J/\psi \phi}$}
\def\D0{D\O}
\def \phis{$\phi_s^{J/\psi \phi}$}
\def\bddec{${B_d^0 \rightarrow J/\psi K^*}$}

\begin{document}

\title{Measurement of the {\boldmath $CP$}-violating  phase  {\boldmath $\phi_s^{J/\psi \phi}$} at \D0}

\author{A. Chandra \em for the \D0 Collaboration}
\affiliation{Department of Physics and Astronomy, Rice University, Houston, TX, USA}

\begin{abstract}
This paper is a report of an updated measurement of the $CP$-violating phase \phis\
and the  decay width difference for the two mass eigenstates $\Delta \Gamma_s$
from flavor-tagged decay ${B_s^0 \rightarrow J/\psi \phi}$.
The 68\% confidence level intervals, including systematic uncertainties, 
are $\phi_s^{J/\psi \phi} =  -0.55 ^{+0.38} _{-0.36}$ and 
$\Delta \Gamma_s =  0.163  ^{+0.065} _{-0.064}$ ps$^{-1}$.
This measurement is in agreement with SM expected value, 
the $p$-value for the Standard Model point is 29.8\%.
The data sample corresponds to an integrated luminosity of 8.0 fb$^{-1}$
accumulated with the D0 detector using $p \overline p$ collisions at $\sqrt{s} = 1.96$ TeV produced
at the Fermilab Tevatron collider.
\end{abstract}

\maketitle

\thispagestyle{fancy}


\section{\label{sec:intro}Introduction}
In the standard model (SM), the light ($L$) and heavy ($H$) mass eigenstates 
of the mixed $B_s^0$ system  are expected to have  sizeable 
mass and decay width differences: $\Delta M_s \equiv M_H - M_L$ and
$\Delta \Gamma_s \equiv \Gamma_L - \Gamma_H$. 
The two mass eigenstates are expected to be almost pure $CP$ eigenstates.  
The $CP$-violating  phase that appears in $b \rightarrow c \overline c  s$ decays,
due to the interference of the decay with and without mixing,
is predicted~\cite{LN2006} to be $\phi_s^{J/\psi \phi} = -2\beta_s = 2\arg[-V_{tb}V^*_{ts}/V_{cb}V^*_{cs}]
 = -0.038 \pm 0.002$, where $V_{ij}$ are elements of the  Cabibbo-Kobayashi-Maskawa 
quark-mixing matrix~\cite{ckm}. 
New phenomena may alter the observed phase~\cite{UTfit}  to $\phi_s^{J/\psi \phi} \equiv -2\beta_s +\phi_s^\Delta$.

The first direct constraint on \phis\ ~\cite{prl07} was derived by analyzing 
${B_s^0 \rightarrow J/\psi \phi}$ decays where the flavor
(i.e., $B^0_s$ or $\overline{B}^0_s$) at the time of production was not 
determined (``tagged'').
It was followed by an improved analysis~\cite{prl08},
based on  2.8 fb$^{-1}$  of integrated luminosity,
that included the  information on the $B_s^0$ flavor  at production.
In that analysis we measured $\Delta \Gamma_s$ and the average lifetime of the \bs\ system, 
$\overline \tau_s =1/\overline \Gamma_s$, where $\overline\Gamma_s \equiv(\Gamma_H+\Gamma_L)/2$.
The $CP$-violating phase \phis\ was also extracted for the first time.
The measurement  correlated two solutions for \phis\ with two corresponding solutions for $\Delta \Gamma_s$.
Improved precision was obtained by refitting the results using additional experimental constraints~\cite{combo}.
Here we present new results from the time-dependent
amplitude analysis of the decay \bsdec\  using a
data sample corresponding to an integrated luminosity 
of 8.0 fb$^{-1}$ collected  with the  D0 detector~\cite{run2det}
at the Fermilab Tevatron Collider.
We measure $\Delta \Gamma_s$;
the average lifetime of the \bs\ system, 
$\overline \tau_s =1/\overline \Gamma_s$, where
$\overline\Gamma_s \equiv(\Gamma_H+\Gamma_L)/2$; and
the {\sl CP}-violating phase \phis.

\section{\label{sec:event} Data Sample and Event Reconstruction}

The analysis presented here is based on data accumulated
between February 2002 and June 2010, corresponds to 8 fb$^{-1}$ of integrated luminosity.

We reconstruct the decay chain \bsdec, $J/\psi \rightarrow \mu ^+ \mu ^-$,
$\phi \rightarrow K^+ K^-$ from  candidate  ($J/\psi,\phi$) pairs
consistent with coming from a common vertex and  having 
an invariant mass in the range $5.37 \pm 0.20$ GeV. 
Events are collected with a mixture of single and dimuon triggers. 
To avoid a bias in the $B_s^0$ lifetime distribution events are rejected
 if they only satisfy triggers that impose a requirement on the track impact parameter
 with respect to the $p\overline{p}$ interaction vertex.

$B^0_s$ candidate events are required to include two opposite-sign muons accompanied by two opposite-sign tracks.
Both muons are required to be detected in the muon chambers inside the toroid magnet 
and at least one of the muons is required to be also detected outside the toroid.
Invariant mass range for muon pairs is $3.096 \pm 0.350$ GeV, 
consistent with $J/\psi$ decay. $J/\psi$ candidates are combined with pairs of oppositely charged tracks 
(assigned the kaon mass) consistent with production at a common vertex, and with an 
invariant mass in the range $1.019 \pm 0.030$ GeV. 
Each of the four final-state tracks is required to have at least one SMT hit.

A kinematic fit under the $B_s^0$ decay hypothesis constrains the dimuon invariant 
mass to the world-average $J/\psi$ mass~\cite{PDG}
and constrains the four-track system to a common vertex.
In events where multiple candidates satisfy these requirements, we select
the candidate with the best decay vertex fit probability.

The primary vertex (PV) is reconstructed using tracks that 
do not originate from the candidate $B_s^0$ decay, 
and apply a constraint to the average beam-spot position in the transverse plane.
We define the signed  decay length of a \bs\ meson, $L^B_{xy}$, 
as the vector pointing from the PV to the decay vertex, projected on the
\bs\ transverse momentum $p_T$.
The proper decay time of a \bs\ candidate is given by
 $t = M_{B_s}\vec L_{xy}^B \cdot \vec{p}/(p_T^2)$
where $M_{B_s}$ is the world-average \bs\ mass~\cite{PDG},
and $\vec p$ is the particle momentum.
The distance in the beam direction between the PV
and the $B_s^0$ vertex is required to be less than 5 cm. 
Approximately 5 million events are accepted after the selection described in this section.

\section{\label{sec:multivar} Background Suppression}

The selection criteria are designed to optimize the measurement of
$\phi_s^{J/\psi \phi}$ and $\Delta \Gamma_s$.
Most of the background is  due to directly 
produced $J/\psi$ mesons accompanied by  tracks arising from 
hadronization.  This ``prompt'' background is distinguished from 
the ``non-prompt'', or ``inclusive $B \rightarrow J/\psi +X$''  background, 
where the $J/\psi$ meson is a product 
of a $b$-hadron decay  while the tracks forming the $\phi$ candidate 
emanate from a multi-body  decay of a $b$ hadron or from hadronization.
Two different event selection approaches are used, one based on a multi-variate
technique, and one based on simple limits on kinematic and event quality parameters.

Three Monte Carlo (MC) samples are used to study background suppression: 
signal, prompt background, and non-prompt background. All three are generated with 
{\sc pythia}~\cite{pythia}. Hadronization is also done in {\sc pythia}, but
all hadrons carrying heavy flavors are passed on to {\sc EvtGen}~\cite{evtgen}  to
model their decays.
   The prompt background MC sample consists of $J/\psi \to \mu^+ \mu^-$ 
decays produced in  $gg \to J/\psi g$, $gg \to J/\psi \gamma$, 
and $g \gamma \to J/\psi g$ processes.
The signal and non-prompt background samples are generated from primary
  $b \bar{b}$  pair production with all $b$ hadrons being produced inclusively and
the $J/\psi$ mesons forced into $\mu^+ \mu^-$  decays. For the signal sample, events
with a \bs\  are selected,  their  decays to $J/\psi \phi$
are implemented without
mixing and with uniform angular distributions, and the \bs\ mean 
lifetime is set to  $\overline \tau_s$ = 1.464 ps. 
There are approximately 10$^6$ events in each background and the signal MC samples.
All events are passed through a full standard chain of {\sc geant}-based~\cite{Geant}
detector software of \D0 simulation.

\section{Multivariate event selection}

To discriminate the signal from background events, we use the TMVA package~\cite{refroot}.
In preliminary studies using MC simulation, the
Boosted Decision Tree (BDT) algorithm was found to demonstrate the best performance.
Since prompt and non-prompt backgrounds have different kinematic behavior, 
we train two discriminants, one for each type of background.
We use  a set of 33 variables for the prompt background
and 35 variables  for the non-prompt background.

\begin{figure}[h!tb]
\begin{center}
 \includegraphics*[width=0.4\textwidth]{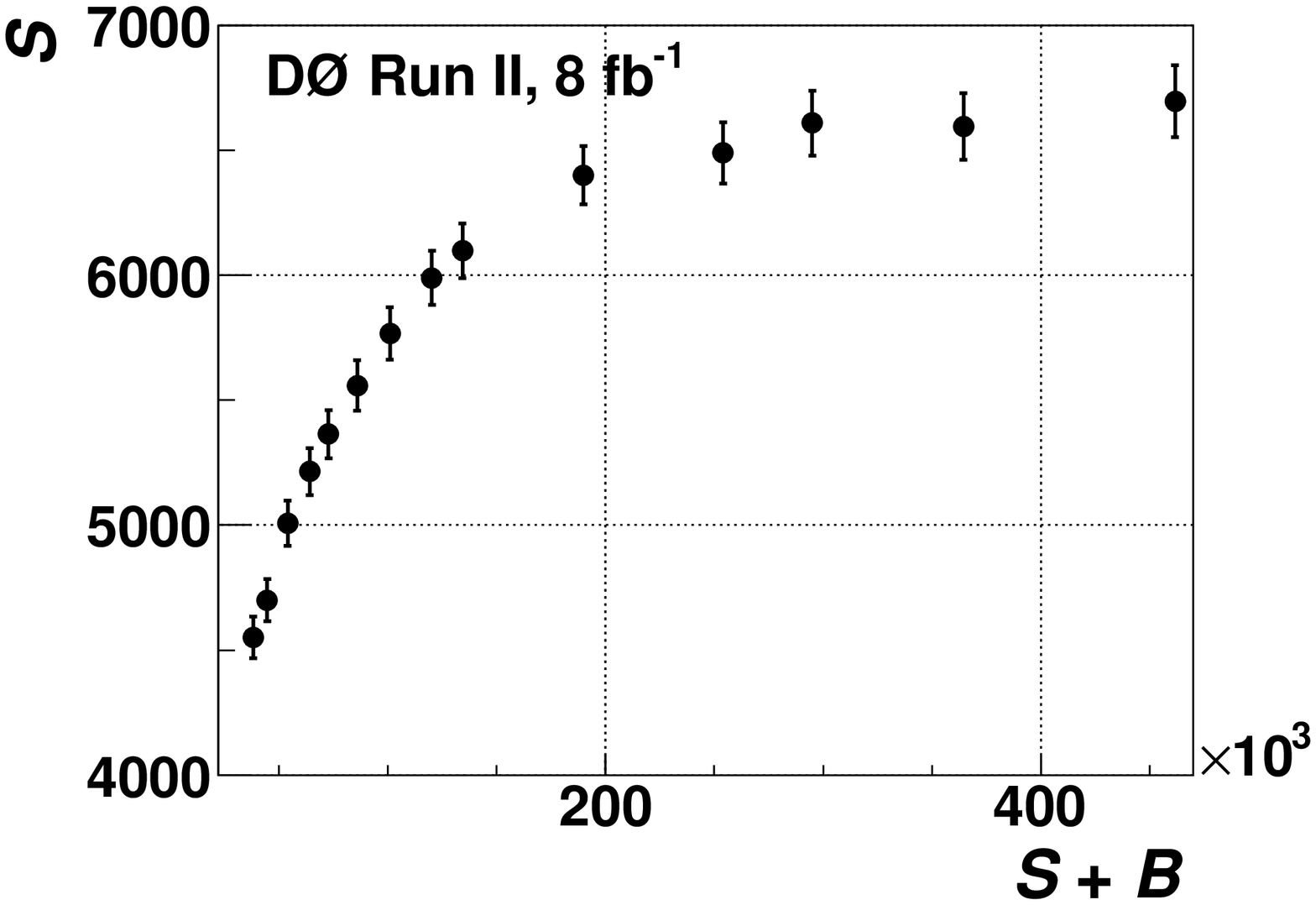}
 \includegraphics*[width=0.4\textwidth]{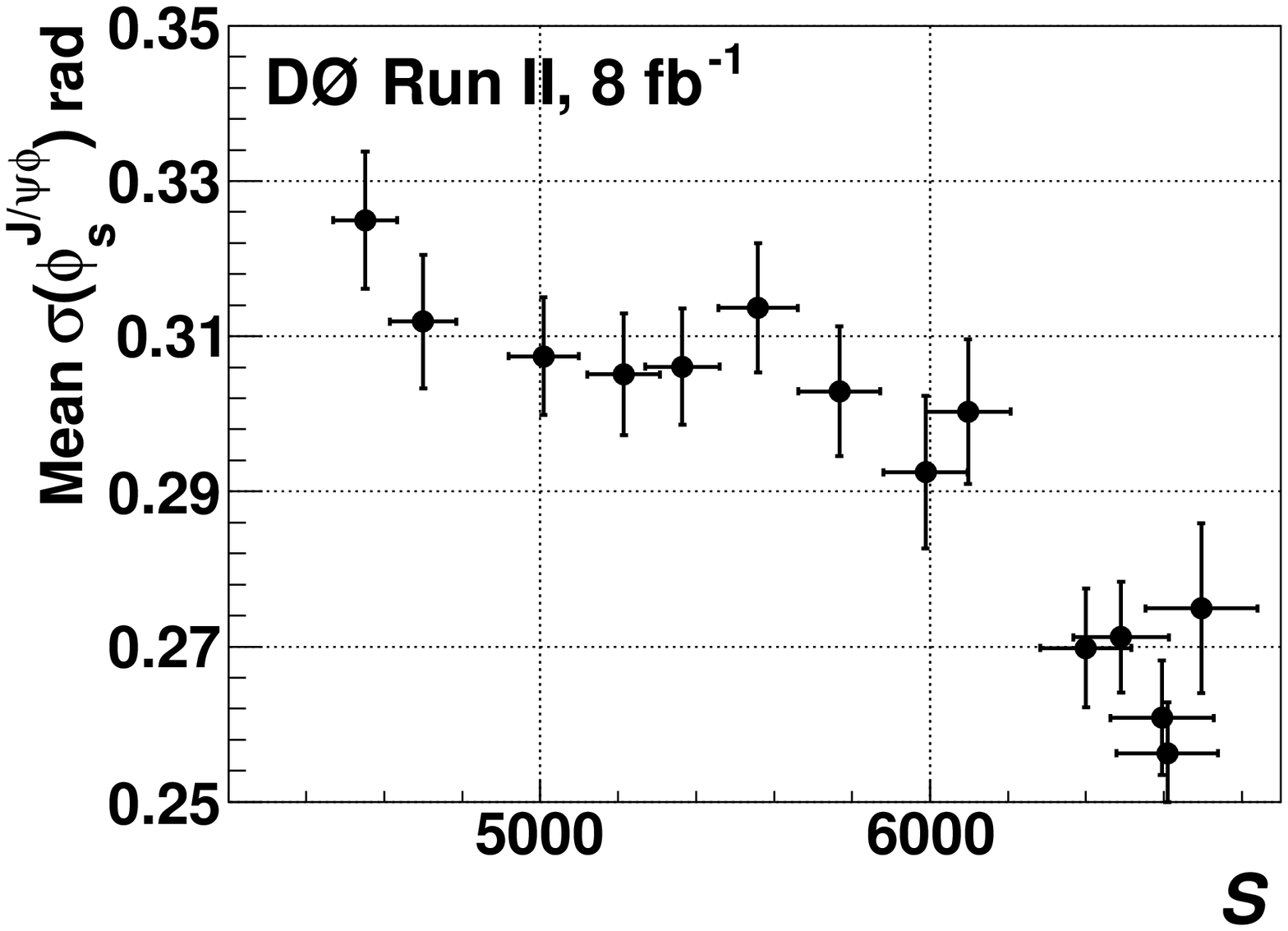}
 \caption{(a) Number of \bsdec\ signal events as a function of the 
total number of events for the 14 criteria sets considered. \ \ \ \
(b) Mean value of $\sigma(\phi_{s})$ as a function of the number of signal events.
}
\label{fig:sigvstot}
\end{center}
\end{figure}

To choose the best set of criteria for the two BDT discriminants, we
start with 14 data samples with signal yeilds ranging from 4000 to 7000
events. For each sample we choose the pair of BDT cuts which gives the
highest significance S/sqrt(S+B), where $S$ ($B$) is the number of signal (background)
events in the data sample. Figure~\ref{fig:sigvstot}(a) shows the 
number of signal events as a function of the total number of events for the 14 points. As the BDT 
criteria are loosened, the total number of events increases by a factor of ten, while the number of 
signal events increases by about 50\%.

The choice of the final cut on the BDT output is based on an ensemble study.  
We perform a maximum-likelihood fit to the event distribution in the 
2-dimensional (2D) space of $B_s^0$ candidate mass and proper time. This 2D fit provides a 
parametrization of the background mass and proper time distribution. We then 
generate pseudo-experiments in the 5D space of $B_s^0$ candidate mass, proper time, and 
three independent angles of decay products, 
using as input the parameters as obtained 
in a preliminary study, and the background from the 2D fit. 
We perform a 5D maximum likelihood fit on the ensembles and compare the  
distributions of the statistical uncertainties of
 $\phi_s^{J/\psi\phi}$ ($\sigma(\phi_s^{J/\psi \phi})$) 
and $\Delta \Gamma_s$ ($\sigma(\Delta \Gamma_s)$) for the different sets of criteria. 
The dependence of the mean values of $\sigma(\phi_s^{J/\psi \phi})$  
on the number of signal events is shown in Figure~\ref{fig:sigvstot}(b).

The mean statistical uncertainties of both $\phi_s^{J/\psi \phi}$ and 
$\Delta \Gamma_s$ systematically decrease with increasing signal, favoring looser 
cuts. The gain in the parameter resolution is slower for the three loosest criteria, while 
the total number of events doubles from about 0.25$\times 10^6$ to 0.5$\times 10^6$.
The fits used for these ensemble tests were simplified, 
therefore the magnitude of the predicted uncertainty is expected to underestimate the
final measured precision. However, the general trends should be valid. Based on 
these results, we choose the sample that contains about 6500 signal events.

We select a second event sample by applying criteria on event quality and kinematic quantities.
We use the consistency of the results obtained for the BDT and for this sample
as a measure of systematic effects related to imperfect modeling of the
detector acceptance and of the selection requirements. The criteria are the same
as in Refs.~\cite{prl07} and \cite{prl08}. 
We refer to this second sample  as the ``Square-cuts'' sample.

\section{\label{sec:flavor} Flavor Tagging }

At the Tevatron, $b$ quarks are mostly produced in $b \overline b$ pairs.
The flavor of the initial state of the $B_s^0$ candidate is determined
by exploiting  properties of particles produced
by the other $b$ hadron  (``opposite-side tagging'', or OST). 
The OST-discriminating variables are based primarily on the presence 
of a  muon or an electron from the semi-leptonic decay of 
the other $b$ hadron produced in the $p\overline p$ interaction. 
If a charged lepton is not found, the algorithm attempts to reconstruct the decay vertex
of the opposite-side $b$ hadron and determine the net charge of particles
forming the vertex. 

The OST algorithm, based on the Likelihood Ratio method,  assigns to each event 
a value of the predicted 
tagging parameter  $d$, in the range [$-1$,1], with $d>0$ tagged as an initial $b$
quark and  $d<0$ tagged as an initial $\overline b$ quark. 
Larger  $|d|$ values correspond to higher tagging confidence.
In events where no tagging information is available  $d$ is set to zero.
 The efficiency $\epsilon$ of the OST, defined as fraction of the number of
candidates with $d \ne 0$, is  18\%.
The OST-discriminating variables and algorithm are described in detail in Ref.~\cite{bflavor}.

The  tagging  dilution ${\cal D}$ is defined as $\cal{D} = N_{\rm cor}-N_{\rm wr}/(N_{\rm cor}+N_{\rm wr})$ \ \ \ 
where $N_{\rm cor}$ ($N_{\rm wr}$) is the number of events with correctly 
(wrongly) identified initial $B$-meson flavor.

The dependence of the tagging dilution on the tagging parameter 
$d$ is calibrated with data for which the flavor ($B$ or $\overline B$) is known.
The dilution calibration is based on four
independent $B_d^0 \to \mu \nu D^{*\pm}$   data samples corresponding 
to different time periods. For each sample we  perform an
analysis of the $B_d^0-\overline{B}_d^0$ oscillations described in Ref.~\cite{Abazov:2006dm}.
We divide the samples in five ranges of the tagging parameter $|d|$,
and for each range we obtain a mean value of the dilution ${\cal |D|}$.
The mixing frequency $\Delta M_d$ is fitted simultaneously and  is found
to be stable and consistent with the world average value.
The measured values of the tagging dilution  $\cal |D|$  for 
the running period of time is parametrized by function:
\begin{equation}
\label{dileq}
{\cal |D|} = \frac{p_0}{(1 + \exp((p_1 - |d|)/p_2))} - \frac{p_0}{(1+\exp(p_1/p_2))}.  
\end{equation}
and the function is fitted to the data.
There is a good agreement in the fits for running period of time, and hence a weighted average is taken.

\section{\label{sec:fitting} Maximum Likelihood Fit}

We perform a six-dimensional (6D) unbinned maximum likelihood fit to the proper decay time
and its uncertainty,
three decay angles characterizing the final state, and the mass of the $B_s^0$ candidate. 
We use events for which the invariant mass of the $K^+K^-$ pair is
within the range 1.01 -- 1.03 GeV.
There are 104683 events in the BDT-based sample and 66455 events in the
Square-cuts sample.
We adopt the formulae and notation of  Ref.~\cite{formulae}.  
The normalized  functional form
of the differential decay rate 
includes an $\cal S$-wave $KK$ contribution in addition to the dominant
$\cal P$-wave $\phi \rightarrow K^+K^-$ decay. 
To model  the distributions of the signal and background we use 
the software library {\sc RooFit}~\cite{refroofit}.

\subsection{Signal model}

The angular distribution of the signal is expressed in the transversity basis.
In the coordinate system of the $J/\psi$ rest frame, 
 where the $\phi$ meson moves in the $x$ direction,
 the $z$  axis is perpendicular to 
the decay plane of $\phi \to K^+ K^-$, and $p_y(K^+)\geq 0$.
The transversity polar and azimuthal angles 
$\theta$ and  $\varphi$ describe the
direction of the positively-charged muon, while $\psi$ is 
the angle between  $\vec p(K^+)$ and  $-\vec{p}(J/\psi)$ 
 in the $\phi$ rest frame.

In the transversity basis, the decay amplitude of the $B_s^0$ and $\overline B_s^0$ mesons 
is decomposed into three independent components corresponding to linear polarization states of 
the vector mesons  $J/\psi$ and  $\phi$, which are polarized either 
longitudinally (0) or transversely to their direction of motion,
and parallel ($\parallel$) or perpendicular ($\perp$) to each other.

The time dependence of amplitudes ${\cal A}_i(t)$ and ${\bar {\cal A}_i}(t)$
($i$ denotes one of  $\{||, \perp, 0\}$),
for $B_s^0$ and $\overline B_s^0$ states to reach the final state $J/\psi$ $\phi$ is:
\begin{eqnarray}
{\cal A}_i(t) &=&    F(t) \left[E_+(t) \pm e^{2i\beta_s} E_-(t)\right] a_i\,,   \nonumber \\
{\bar {\cal A}_i}(t) &=& F(t) \left[\pm E_+(t) + e^{-2i\beta_s} E_-(t)\right] a_i\,
\label{eqn:finalAmp}
\end{eqnarray}
where 
\begin{eqnarray}
F(t)  &=& \frac{e^{-\Gamma_s t /2}}{\sqrt{\tau_H + \tau_L \pm \cos{2\beta_s}\left(\tau_L-\tau_H\right)}} \,,
\end{eqnarray}
and $\tau_H $ and $ \tau_L$ are the lifetimes of the heavy and light $B_s^0$ eigenstates.

In the above equations the upper sign indicates a {\sl CP}-even final state, the lower sign indicates a {\sl CP}-odd final state,
\begin{equation}
E_{\pm}(t) \equiv  \frac{1}{2}\left[e^{\left(\frac{-\Delta\Gamma_s}{4} +
i\frac{\Delta M_s}{2}\right)t} \pm e^{-\left(\frac{-\Delta\Gamma_s}{4} + i\frac{\Delta M_s}{2}\right)t}\right],
\label{eqn:functionDef}
\end{equation}
and the amplitude parameters  $a_i$ give the time-integrated decay rate to each of the polarization states,
$|a_i|^2$,   satisfying: \ \ \ \ \ $\sum_i {|a_i|^2}$ = 1. \ \ \ The  normalized probability density functions 
$P_{B_z}$ ($X_z$ denotes one more same equation replaced by $\bar {X_z}$) 
\begin{eqnarray} 
P_{B_z}(\theta, \varphi, \psi, t) = \frac{9}{16\pi} |{\bf A_z}(t)\times \hat{n}|^2,  \nonumber \\
\label{eqn:finalAngle}
\end{eqnarray}
where $\hat n$ is the muon momentum direction in the $J/\psi$ rest frame,
\begin{equation}
\hat{n} = \left(\sin{\theta}\cos{\varphi}, \sin{\theta}\sin{\varphi}, \cos{\theta} \right),
\label{eqn:nhat}
\end{equation}
and  ${\bf A}(t)$ and ${\bf {\bar A}}(t)$ are complex vector functions of time defined as

\begin{eqnarray}
{\bf A_z}(t)=\left({\mathcal (A_{z})}_0(t)\cos{\psi}, -\frac{{\mathcal (A_{z})}_\parallel(t)\sin{\psi}}{\sqrt{2}}, i\frac{{\mathcal (A_{z})}_\perp(t)\sin{\psi}}{\sqrt{2}}\right),  \nonumber \\
\label{eqn:fixedAngle}
\end{eqnarray}

The values of ${\cal A}_i(t)$ at $t=0$ are denoted as $A_i$. They are related to the parameters $a$ by 
\begin{eqnarray}
|A_{i}|^2 = \frac{|a_{i}|^2y}{1+(y-1)|a_{\perp}|^2},   \nonumber \\
\end{eqnarray}
where $y \equiv (1-z)/(1+z)$ and $z\equiv \cos{2\beta_s}\Delta\Gamma_s/(2\overline \Gamma_s)$.
By convention,  the phase of $A_0$ is set to zero and the phases 
of the other two amplitudes are denoted by $\delta_{||}$ and  $\delta_\perp$. 

For a given event, the decay rate is the sum of the 
functions $P_B$ and $P_{\bar B}$  weighted  by the flavor tagging
dilution factors $(1+ {\cal D})/2$ and $(1-{\cal D})/2$, respectively.

The contribution from the decay to $J/\psi K^+K^-$ with the kaons in an $\cal S$ wave
is expressed in terms of the $\cal S$-wave fraction $F_S$ and a phase $\delta_s$.
The squared sum of the $\cal P$ and $\cal S$ waves is integrated over the $KK$ mass.
For the $\cal P$ wave, we assume the non-relativistic Breit-Wigner model

\begin{equation}
  g(M(KK)) = \sqrt{\frac {\Gamma_{\phi}/2}{\Delta M(KK)}} \cdot \frac{1}{M(KK) - M_{\phi} + i \Gamma_{\phi}/2}
\label{eqn:gDef}
\end{equation}
with the  $\phi$ meson mass $M_{\phi} = 1.019$ GeV and width $\Gamma_{\phi} =4.26$ MeV~\cite{PDG},
and with $\Delta M(KK) = 0.02$ GeV.

For the $\cal S$-wave component, we assume
a uniform distribution in the range $1.01 < M(KK)<1.03$ GeV.
In the case of the BDT selection, it is modified by a $KK$-mass dependent
factor corresponding to the BDT selection efficiency.
We constrain the
oscillation frequency to $\Delta M_s = 17.77 \pm 0.12$ ps$^{-1}$, 
as measured in Ref.~\cite{dms}. 
Table~\ref{sigpar} lists all physics parameters used in the fit.

\begin{table}[htbp]
\begin{tabular}{cc}
\hline
\hline
                  Parameter           &    Definition \tabularnewline
\hline

$|A_0|^2$                       &  $\cal P$-wave longitudinal  amplitude squared, at $t=0$ \tabularnewline
$A1$        & $|A_{\|}|^2/(1-|A_0|^2)$     \tabularnewline
$\overline{\tau}_s$ (ps)        & $B^0_s$ mean lifetime  \tabularnewline
$\Delta\Gamma_s$ (ps$^{-1}$)    &  Heavy-light decay width difference   \tabularnewline
$F_S$                           &  $K^+K^-$ $\cal S$-wave fraction \tabularnewline
$\beta_s$                       &  {\sl CP}-violating phase  ( $\equiv -\phi_s^{J/\psi \phi}/2$)  \tabularnewline
$\delta_{\|}$                   &  $\arg(A_{\|}/A_0)$     \tabularnewline
$\delta_{\perp}$                &  $\arg(A_{\perp}/A_0)$  \tabularnewline
$\delta_s$                      &  $\arg(A_{s}/A_0)$      \tabularnewline
\hline
\hline
\end{tabular}
\caption {Definition of nine real measurables for the decay \bsdec\ used in the
Maximum Likelihood fitting.
}
\label{sigpar}
\end{table}

For the signal mass distribution we use a Gaussian function with
a free  mean value, width, and normalization.
The function describing the signal rate in the 6D
space is invariant under the combined transformation 
$\beta_s \rightarrow \pi/2 - \beta_s$,
$\Delta \Gamma_s  \rightarrow  -\Delta \Gamma_s$, 
$\delta_{\|}  \rightarrow  2\pi - \delta_{\|} $, 
$\delta_{\perp}  \rightarrow  \pi - \delta_{\perp} $, and
$\delta_s  \rightarrow  \pi - \delta_s$.
In addition, with a limited flavor-tagging power,
there is an approximate symmetry around $\beta_s =0$ 
for a given sign of $\Delta \Gamma_s$.

We correct the signal decay rate by 
a detector acceptance factor $\epsilon(\psi, \theta, \varphi)$ 
parametrized by  coefficients 
of expansion in Legendre polynomials
$P_k(\psi)$ and real harmonics $Y_{lm}(\theta,\varphi)$. The coefficients
are obtained from Monte Carlo simulation.

\subsection{Background model}

The proper decay time distribution  of the  background 
is described by a sum of
a prompt component, modeled as 
a Gaussian function centered at zero, and a non-prompt component.
The non-prompt component  is modeled as a superposition of one 
exponential decay for $t<0$ 
and two exponential decays for $t>0$, with free slopes and normalizations.
The lifetime resolution is modeled by an exponential convoluted with 
a Gaussian function,
with two separate parameters for prompt and non-prompt background.
To allow for the possibility of the lifetime uncertainty to be 
systematically
underestimated, we introduce a free scale factor.

The mass distributions of the  two components of background
are parametrized by low-order polynomials: a linear function for the
prompt background and a quadratic function for the non-prompt background.
The angular distribution of background is parametrized
by Legendre and real harmonics expansion coefficients. A separate set of 
expansion coefficients $c^k_{lm}$ and  $c^k_{lm}$, with
$k=0$ or $2$ and $l=0,1,2$, is used for the prompt and non-prompt background.
A preliminary fit is first performed with all $17\times2$ parameters allowed to vary. 
In subsequent fits
those that converge at values  within two standard deviations of zero are set to zero.
Nine free parameters remain, five for non-prompt background:
$c^0_{1-1}$,   $c^0_{20}$,  $c^0_{22}$, $c^2_{00}$, and
$c^2_{22}$, 
and four for prompt background:
$c^0_{1-1}$,   $c^0_{20}$,  $c^0_{22}$,  and  $c^2_{2-1}$.
All background parameters described above are varied simultaneously with
physics  parameters. 
In total, there are 36 parameters used in the fit. In addition
to the  nine physics parameters defined
in Table~\ref{sigpar}, they are: signal yield, mean mass and width, 
non-prompt background contribution, six non-prompt background lifetime
parameters, four background time resolution parameters, one 
time resolution scale factor, three 
background mass distribution parameters, and nine parameters describing
background angular distributions.

\subsection{\label{sec:syst} Systematic uncertainties}

There are several possible sources of systematic uncertainty in
the measurements. 
These uncertainties are estimated for:

\begin{itemize}

\item {\bf Flavor tagging}:
The nominal calibration of the
flavor tagging dilution is determined as a weighted average of four
samples separated by the running period. As an alternative,
we  use two separate calibration parameters, for the same running period.
We also alter the nominal parameters by their uncertainties.

\item {\bf Proper decay time resolution}:
Fit results can be affected by the uncertainty of 
the assumed proper decay time resolution function. To assess the effect,
we have used two alternative parameterizations obtained by random
sampling of the resolution function.

\item {\bf Detector acceptance}:
The effects of imperfect modeling of the
detector acceptance and of the
selection requirements are estimated by investigating the
consistency of the fit results
for the sample based on the BDT selection and on the Square-cuts selection.
Although the overlap between the two samples is
70\%, and some statistical differences are
expected, we interpret the differences in the results
as a measure of systematic effects.

\item {\bf {\boldmath $M(KK)$} resolution:}
The limited $M(KK)$ resolution may affect the results
of the analysis, especially the phases and the $\cal S$-wave fraction $F_S$, 
through the dependence of the $\cal S - \cal P$ interference term
on the $\cal P$-wave mass model. We repeat the fits using this altered 
$\phi(1020)$ propagator as a measure of the sensitivity to
the $M(KK)$ resolution.

\end{itemize}

The differences between the best-fit values and the alternative fit values
provide a measure of systematic effects. For the best estimate of the C.L. ranges for all the measured physics 
quantities, we conduct Markov Chain Monte Carlo (MCMC) technique described in the next section.

\section{\label{sec:mcmc} Confidence intervals from MCMC studies}

The maximum likelihood fit provides the best values of all free parameters,
including the signal observables and background model parameters,
their statistical uncertainties and  their full correlation matrix. 

In addition to the free parameters determined in the fit,
the model depends on a number of external constants whose inherent
uncertainties are not taken into account in a given fit.
Ideally, effects of uncertainties of external constants,
such as time resolution parameters, flavor tagging dilution
calibration, or detector acceptance, should be included in the model
by introducing the appropriate parametrized probability density functions
and allowing the parameters to vary. Such a procedure of proper
integrating over the external parameter space would greatly increase
the number of free parameters and would be prohibitive.
Therefore, as a trade-off, we apply a random sampling
of external parameter values within their uncertainties,
we perform the analysis for thus created ``alternative universes'',
and we  average the results. To do the averaging in the multidimensional
space, taking into account non-Gaussian parameter distributions
and correlations, we use the MCMC technique.

The MCMC  technique uses the
Metropolis-Hastings algorithm \cite{MetHast} to generate a random sample
proportional to  a given probability distribution.
The algorithm generates  a sequence of  ``states'', a Markov chain,
 in which each state depends
only on the previous state. 

To generate a Markov chain for a given maximum likelihood fit 
result, we start from the best-fit point $x$.  We randomly generate 
a point $x'$ according to the multivariate normal  distribution
$\exp(-(x'-x)\cdot \Sigma \cdot (x'-x)/2)$, where
$\Sigma$ is the covariance matrix. The new point
is accepted if ${\cal L}(x')/{\cal L}(x)>1$,
otherwise it is accepted with the probability 
${\cal L}(x')/{\cal L}(x)$.
The process is continued
until a desired number of states is achieved.
To avoid a bias due to the choice of the initial state,
we discard the early states which may ``remember'' the initial state.
Our studies show  that the initial state is ``forgotten'' after approximately
50 steps. We discard the first 100 states in each chain.

While we do not use any external numerical constraints on the polarization
amplitudes, we note that the best-fit values of their magnitudes
and phases are consistent with  those measured in the $U(3)$-flavor related 
decay \bddec\ \cite{PDG}, up to the
sign ambiguities. Ref.~\cite{gr} predicts
that the phases of the polarization amplitudes in the two decay processes 
should agree within approximately 0.17 radians.
For $\delta_{\perp}$, our measurement gives equivalent solutions
near $\pi$ and near zero, with only the former being in agreement with
the value of $2.91 \pm 0.06$ measured for \bddec\ by $B$ factories.
Therefore, in the following we limit the range of  $\delta_{\perp}$
to $\cos \delta_{\perp}<0$.

\subsection{Results}

The fit assigns $5598 \pm 113$  ($5050\pm105$) events to the signal
for the BDT (Square-cuts) sample. A single fit does not provide  meaningful 
point estimates and uncertainties for the four phase parameters.
Their estimates are obtained using the MCMC technique.

\begin{figure}[h!tb]
\begin{center}
 \includegraphics*[width=0.4\textwidth]{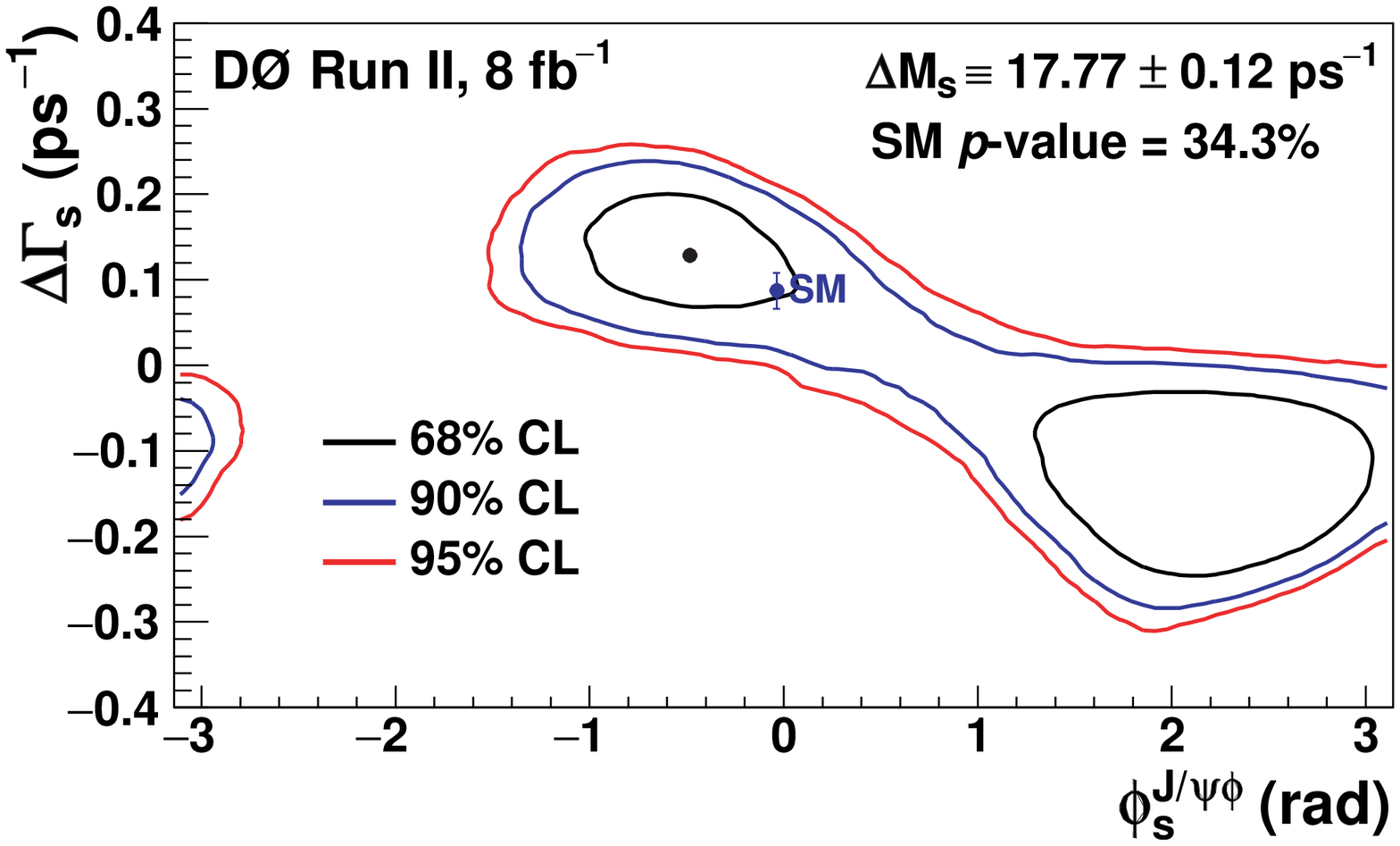}
 \includegraphics*[width=0.4\textwidth]{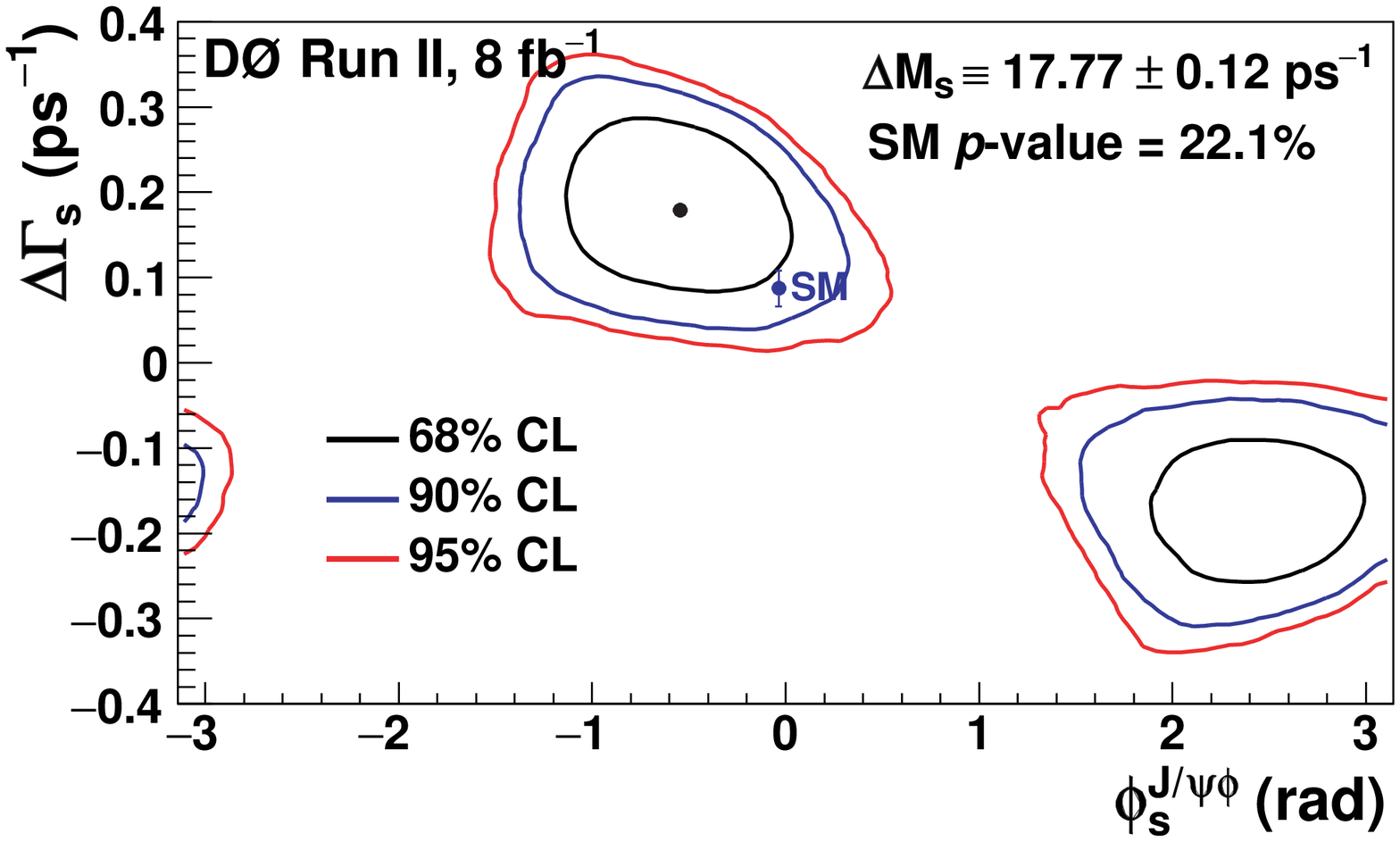}
 \caption{Two-dimensional 68\%, 90\% and  95\% C.L. contour for BDT and Square cuts selection.
The standard model expectation is indicated as a point with an error.
  }
\label{fig:contour_bdtsq}
\end{center}
\end{figure}

Figure~\ref{fig:contour_bdtsq} shows 68\%, 90\% and 95\% C.L. contours
in the   (\phis,$\Delta \Gamma_s$) plane for the BDT-based  
and for the Square-cuts samples.
The point estimates of physics parameters are obtained from
one-dimensional projections.
The minimal range containing 68\% 
of the area of the probability density function defines
the one standard deviation  C.L. interval for each parameter,
while the most probable value defines the central value.

The one-dimensional estimates of physics parameters for the BDT-cuts and Square-cuts sample are 
shown in Table \ref{tab:WithSysRoofitRes}.

\begin{widetext} 
\begin{table}[htbp]
\begin{tabular}{cccc}
\hline
\hline
     Parameter      &  BDT-cut sample  & Square-Cut sample & Final Result \tabularnewline
\hline

$\overline{\tau}_s$ (ps)        & $1.426^{+0.035}_{-0.032}$      & ~~$1.444 ^{+0.041} _{-0.033}$  & ~~$1.443 ^{+0.038}  _{-0.035}$  \tabularnewline
$\Delta\Gamma_s$ (ps$^{-1}$)    & $0.129 ^{+0.076} _{-0.053}$    & ~~$0.179 ^{+0.059} _{-0.060}$  & ~~$0.163  ^{+0.065} _{-0.064}$  \tabularnewline
$\phi_s^{J/\psi \phi}$          & $-0.49 ^{+0.48} _{-0.40}$      & ~~$-0.56 ^{+0.36} _{-0.32}$    & ~~$-0.55 ^{+0.38} _{-0.36}$    \tabularnewline
$|A_0|^2$                       & $0.552 ^{+0.016} _{-0.017}$    & ~~$0.565 \pm 0.017$            & ~~$0.558 ^{+0.017} _{-0.019}$  \tabularnewline
$|A_{\parallel}|^2$             & $0.219 ^{+0.020} _{-0.021}$    & ~~$0.249 ^{+0.021} _{-0.022}$  & ~~$0.231  ^{+0.024} _{-0.030}$  \tabularnewline
$\delta_{\|}$                   & $3.15 \pm 0.27$                & ~~$3.15 \pm 0.19$              & ~~$3.15 \pm 0.22$              \tabularnewline
$\cos(\delta_{\perp} -\delta_s)$& $-0.06 \pm 0.24$               & ~~$-0.20 ^{+0.26}_{-0.27}$     & ~~$-0.11 ^{+0.27} _{-0.25}$     \tabularnewline
$F_S$                           & $0.146 \pm 0.035$              & ~~$0.173 \pm 0.036$            & ~~$0.173 \pm 0.036$            \tabularnewline
\hline
\hline
\end{tabular}
\caption {The one-dimensional estimates of physics parameters for the BDT-cuts sample, Square-cuts sample
and final result values with systematic error.}
\label{tab:WithSysRoofitRes}
\end{table}
\end{widetext}

\begin{figure}[htbp]
\begin{center}
 \includegraphics*[width=0.6\textwidth]{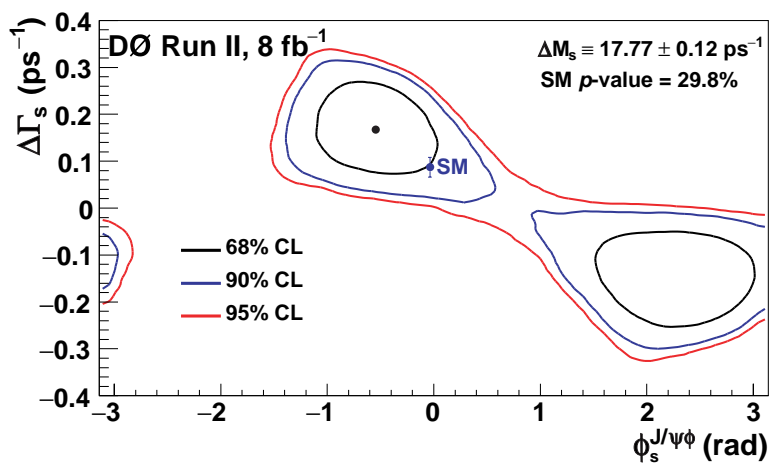}
\caption{Two-dimensional 68\%, 90\% and  95\% C.L. contours including systematic uncertainties.
The standard model expectation is indicated as a point with an error. }
\label{fig:contour_final}
\end{center}
\end{figure}

To obtain the final C.L. ranges for physics parameters, 
 we combine all eight  MCMC chains, effectively averaging the
probability density functions of the results of the fits to the
BDT- and Square-cuts samples.
Figure~\ref{fig:contour_final} shows 68\%, 90\% and 95\% C.L. contours
in the   (\phis,$\Delta \Gamma_s$) plane.
The $p$-value for the SM point~\cite{ln2011}
 ($\phi_s^{J/\psi \phi}, \Delta \Gamma_s)  = (-0.038, 0.087$ ps$^{-1}$)
is 29.8\%.

\section{\label{sec:conclusions}Summary and Discussion}

We have presented  a time-dependent  angular analysis of the decay process \bsdec.
We measure $B_s^0$  mixing parameters,  average lifetime, and decay amplitudes.
In addition, we  measure the amplitudes and phases of the polarization amplitudes.
We also measure the level of the  $KK$ $\cal S$-wave contamination in the mass 
range $1.01$ -- $1.03$ GeV, $F_S$.
The final result values for the 68\% C.L. intervals, including systematic uncertainties, with the
oscillation frequency constrained to  $\Delta M_s = 17.77 \pm 0.12$ ps$^{-1}$, are shown in the 
last column of Table \ref{tab:WithSysRoofitRes}.
The $p$-value for the SM point ($\phi_s^{J/\psi \phi}, \Delta \Gamma_s)  = (-0.038, 0.087$ ps$^{-1}$) is 29.8\%. \\

We thank the staffs at Fermilab and collaborating institutions,
and acknowledge support from the DOE and NSF (USA);
CEA and CNRS/IN2P3 (France); FASI, Rosatom and RFBR (Russia);
CNPq, FAPERJ, FAPESP and FUNDUNESP (Brazil); DAE and DST (India);
Colciencias (Colombia); CONACyT (Mexico); KRF and KOSEF (Korea);
CONICET and UBACyT (Argentina); FOM (The Netherlands);
STFC and the Royal Society (United Kingdom);
MSMT and GACR (Czech Republic); CRC Program and NSERC (Canada);
BMBF and DFG (Germany); SFI (Ireland);
The Swedish Research Council (Sweden); and CAS and CNSF (China).


\end{document}